\documentclass[sigconf]{acmart}

\AtBeginDocument{%
  \providecommand\BibTeX{{%
    \normalfont B\kern-0.5em{\scshape i\kern-0.25em b}\kern-0.8em\TeX}}}

\setcopyright{acmcopyright}
\copyrightyear{2022}
\acmYear{2022}
\setcopyright{acmlicensed}\acmConference[WebSci '22]{14th ACM Web Science Conference 2022}{June 26--29, 2022}{Barcelona, Spain}
\acmBooktitle{14th ACM Web Science Conference 2022 (WebSci '22), June 26--29, 2022, Barcelona, Spain}
\acmPrice{15.00}
\acmDOI{10.1145/3501247.3531570}
\acmISBN{978-1-4503-9191-7/22/06}

\usepackage{amsmath}
\usepackage{amsfonts}

\usepackage[most]{tcolorbox}

\usepackage[utf8]{inputenc}
\usepackage{tikz,lipsum}

\usepackage{subfigure}
\usepackage{caption}
\usepackage{multirow}

\usepackage{cleveref}
\crefname{figure}{Fig.}{Figs.}
\Crefname{figure}{Figure}{Figures}
\Crefname{section}{Section}{Sections}
\crefname{section}{Sect.}{Sect.}
\crefname{subsection}{Sect.}{Sect.}
\Crefname{subsection}{Section}{Sections}
\crefname{subsubsection}{Sect.}{Sect.}
\Crefname{subsubsection}{Section}{Sections}
\crefname{table}{Table}{Tables}
\Crefname{table}{Table}{Tables}
\crefname{exmp}{Example}{Examples}
\Crefname{exmp}{Example}{Examples}

\newcommand{\mbmc}{\texttt{\#MyBodyMyChoice}}

\newcommand{\stream}[1]{S_{#1}}

\newcommand{\covid}{Covid-19}
\begin{document}

\title{The Drift of \#MyBodyMyChoice\ Discourse on Twitter}

\author{Cristina Menghini}
\affiliation{%
 \institution{Brown University}
 \country{USA}
}
\email{cristina\_menghini@brown.edu}

\author{Justin Uhr}
\affiliation{%
 \institution{Brown University}
 \country{USA}
}
\email{justin\_uhr@brown.edu}

\author{Shahrzad Haddadan}
\affiliation{%
 \institution{Brown University}
 \country{USA}
}
\email{shahrzad\_haddadan@brown.edu}

\author{Ashley Champagne}
\affiliation{%
 \institution{Brown University}
 \country{USA}
}
\email{ashley\_champagne@brown.edu}

\author{Bj\"orn Sandstede}
\affiliation{%
 \institution{Brown University}
 \country{USA}
}
\email{bjorn\_sandstede@brown.edu}

\author{Sohini Ramachandran}
\affiliation{%
 \institution{Brown University}
 \country{USA}
}
\email{sohini\_ramachandran@brown.edu}








\renewcommand{\shortauthors}{Menghini et al.}

\begin{abstract}

\mbmc\ is a well-known hashtag originally created to advocate for women’s rights, often used in discourse about abortion and bodily autonomy. 
The \covid\ outbreak prompted governments to take containment measures such as vaccination campaigns and mask mandates.
Population groups opposed to such measures started to use the slogan ``My Body My Choice'' to claim their bodily autonomy.
In this paper, we investigate whether the discourse around the hashtag \mbmc\ on Twitter changed its usage after the \covid\ outbreak.
We observe that the conversation around the hashtag changed in two ways. First, semantically, the hashtag \mbmc\ drifted towards conversations around \covid, especially in messages opposed to containment measures. Second, while before the pandemic users used to share content produced by experts and authorities, after \covid\ the users' attention has shifted towards individuals.

\end{abstract}





\maketitle

\section{Introduction}
Many factors impact what and how users write on online social networks (OSNs).
These factors include platform-level changes such as when Twitter changed the number of characters per tweet from 140 to 240 in 2017  \cite{gligoric2018constraints}, external events such as a presidential election \cite{enli2017twitter,yaqub2020location}, social movements such as the Arab spring or the Black Lives Matter movement \cite{starbird2012will,ince2017social}, or crises such as earthquakes or terrorist attacks \cite{mendoza2010twitter,cassa2013twitter}. In this paper, we focus on how users adopt hashtags for a different purpose than originally intended or \emph{hijack} hashtags. Understanding how users hijack hashtags is important for tracking and predicting social movements within public discourse.

\begin{figure}[t]
\centering
\includegraphics[width=.8\columnwidth]{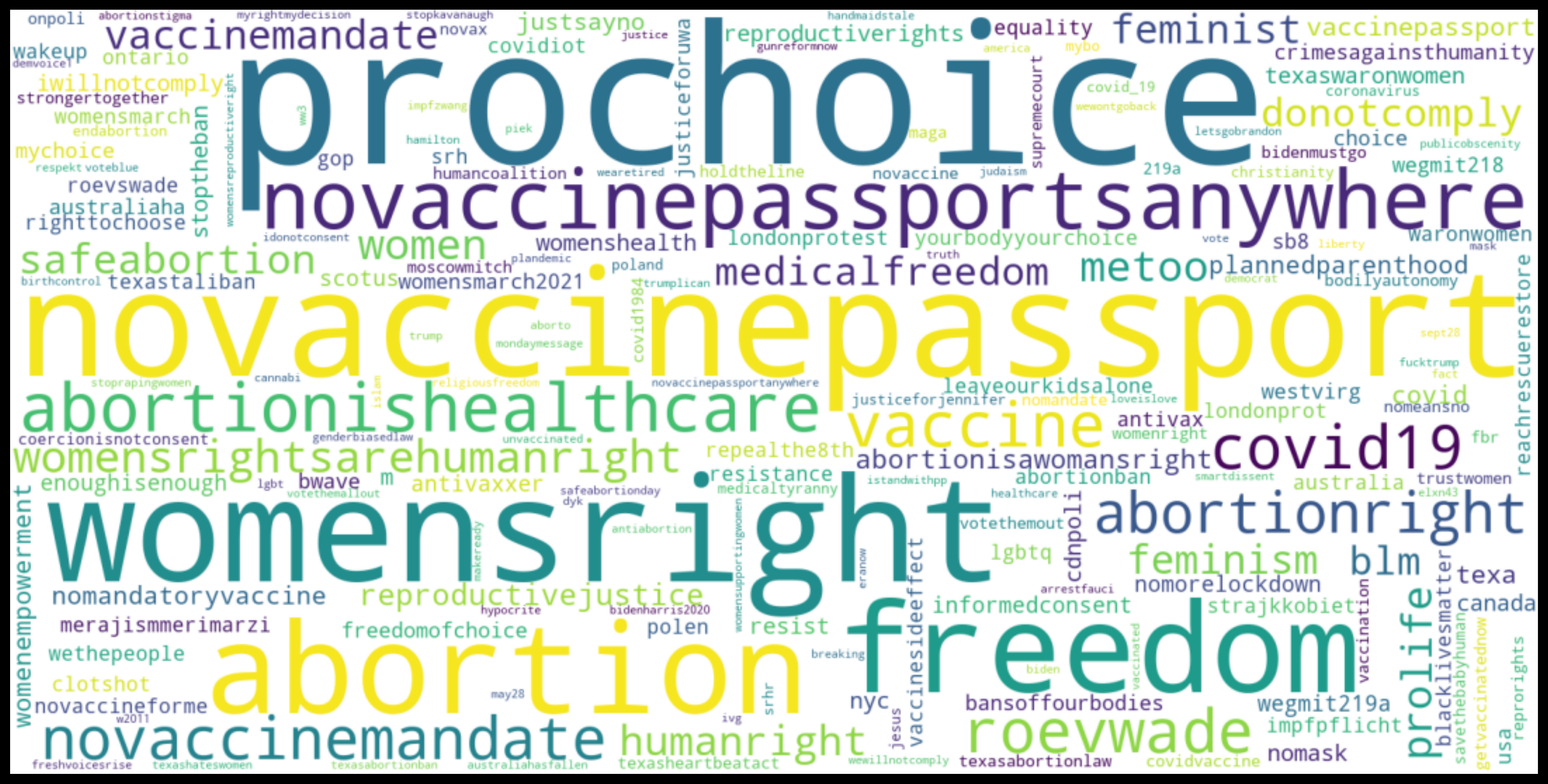}
\caption{Word cloud of the hashtags that co-occur with \mbmc\ in the entire stream of tweets from 1-Jan-2018 to 20-Dec-2021. }
\label{fig:al_word_cloud}
\end{figure}

The global health crisis resulting from the \covid\ pandemic has impacted many aspects of daily life, ranging from health to economics, and has changed in certain ways how users behave on OSNs.
Researchers have shown the role of OSNs in the distribution of false news \cite{vosoughi2018FlasenewsRT}, in part leading to our current \emph{infodemic}. Many researchers have studied the effect of \covid\ on Twitter, which is one of the most potent online platforms for distribution of information. Some research shows that during the \covid\ pandemic, user behavior changed on Twitter and other platforms, such as Wikipedia~\cite{ribeiro2021sudden,ruprechter2021volunteer}.~\citet{gligoric2020experts} shows that Twitter users' attention shifted towards accounts associated to  healthcare, science, government and politics  as opposed to  accounts related to religion or sports.
As the impact of OSNs for spreading hate speech and false news have raised concern, it is vital that online social network administrators concentrate their efforts on excluding organized hate speech or the spread of false news.
A drift in the usage of a hashtag can make it more difficult to identify false news or hate speech \cite{darius2019hashjacking,darius2021far}. 

On Twitter, discussions and topics are organized by usage of hashtags.
When users adopt a hashtag that already exists for another purpose than originally intended, this is often called ``hashjacking'' or referred to as hashtag drift or co-option. Researchers have studied hashjackings through a \emph{linguistic perspective}, by analyzing the tweet itself, and through a focus on \emph{user behaviours}, by researching how user groups interact with tweets \cite{hashjacktopic,detectingHashjackingtwitter,darius2019hashjacking,isisHijack,meetoochange} (see \Cref{sec:tax} for a full discussion).

\smallskip 

In this work, we examine the hashtag \mbmc\ (\Cref{fig:al_word_cloud}) to see how the content of the tweets associated with it evolved during the pandemic. The hashtag originated to support women's rights around abortion and bodily autonomy, but during the pandemic a separate set of users utilized the hashtag to communicate how vaccine mandates related to \covid\ restricted their bodily autonomy. 

Our study shows that the hashtag \mbmc\ can no longer be uniquely associated to the original discourse centered around women's rights.
Indeed, after the \covid\ outbreak there is an increasing number of monthly tweets, up to 75\%, related to \emph{covid \& vaccines}.
Although the discourse about \mbmc\ branches out into several directions, the characterization of users interactions on the corpora of tweets related to \emph{women's rights \& abortion} and \emph{covid \& vaccines} suggests that the topic drift was not intentionally planned by a group of users. 
In~\Cref{sec:tax}, we provide a taxonomy of hashjackings including (1) semantic hashjacking; (2)
hashjacking with polarization intents; (3) hashjacking to capitalize on the success of the original hashtag.
We associate the case \mbmc\ to a semantic hashjacking, which involves a natural topic shift.

\begin{figure*}[ht]
    \centering
    \includegraphics[width=.68\columnwidth]{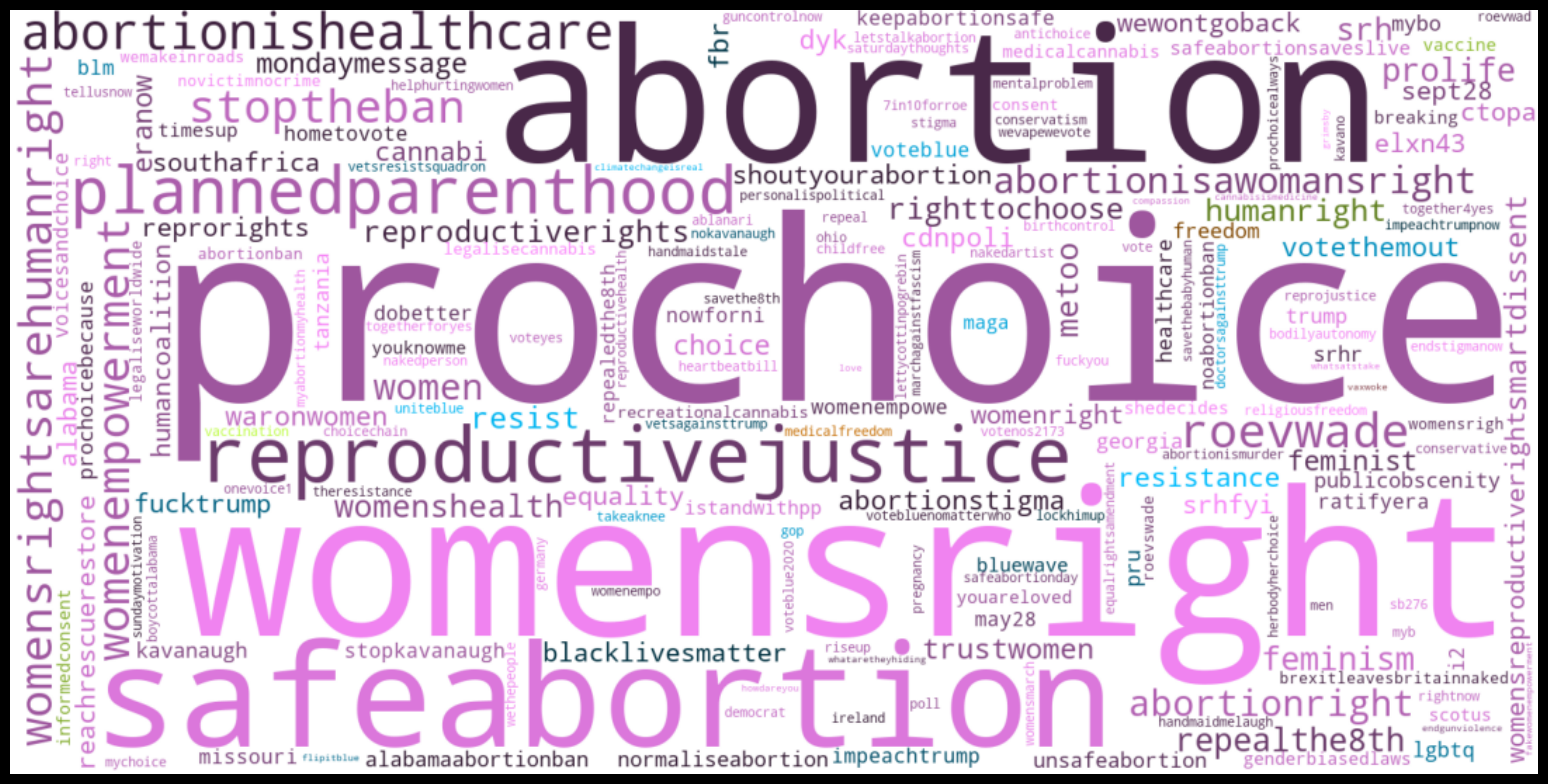}
    \includegraphics[width=.68\columnwidth]{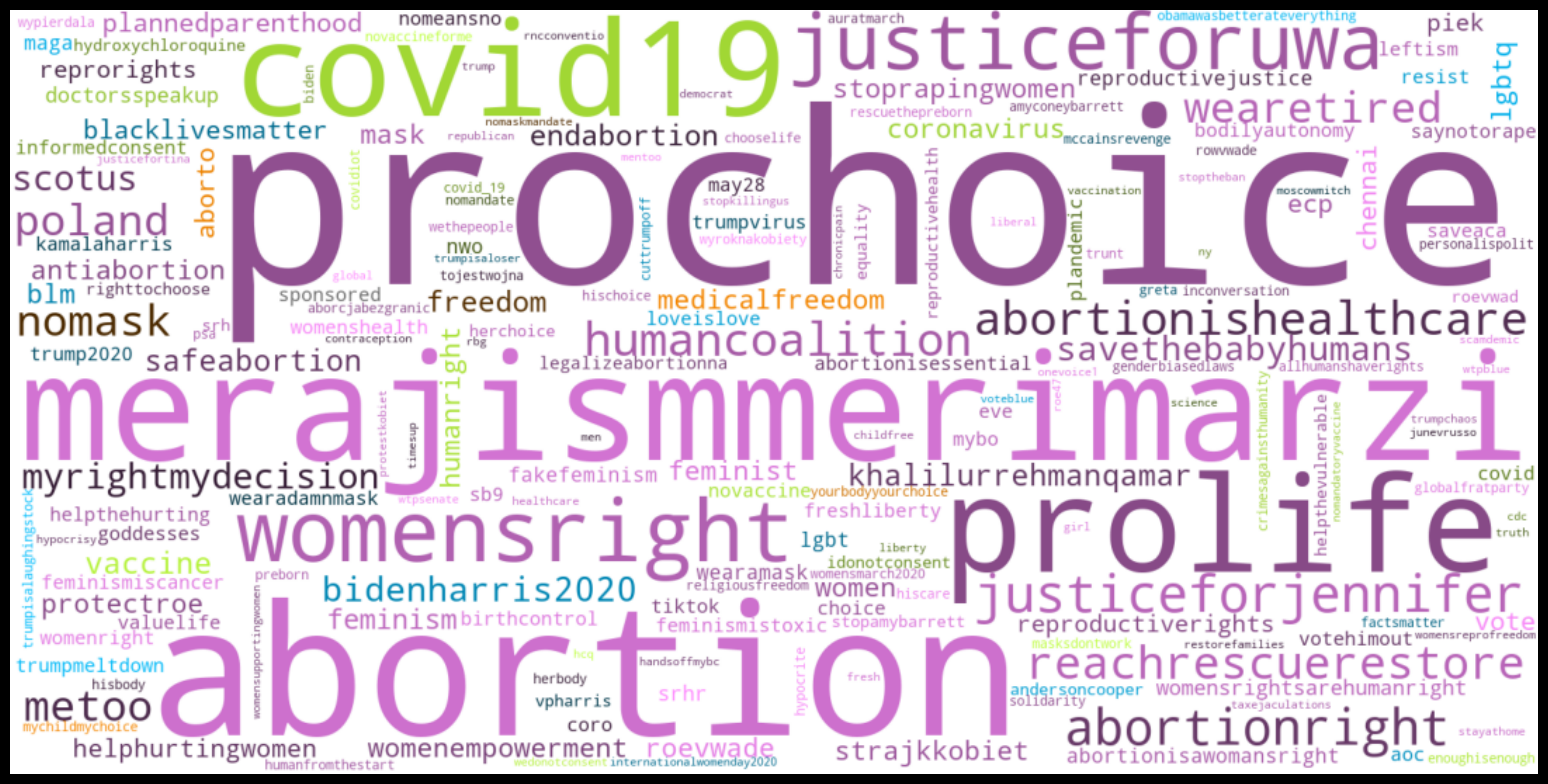}
    \includegraphics[width=.68\columnwidth]{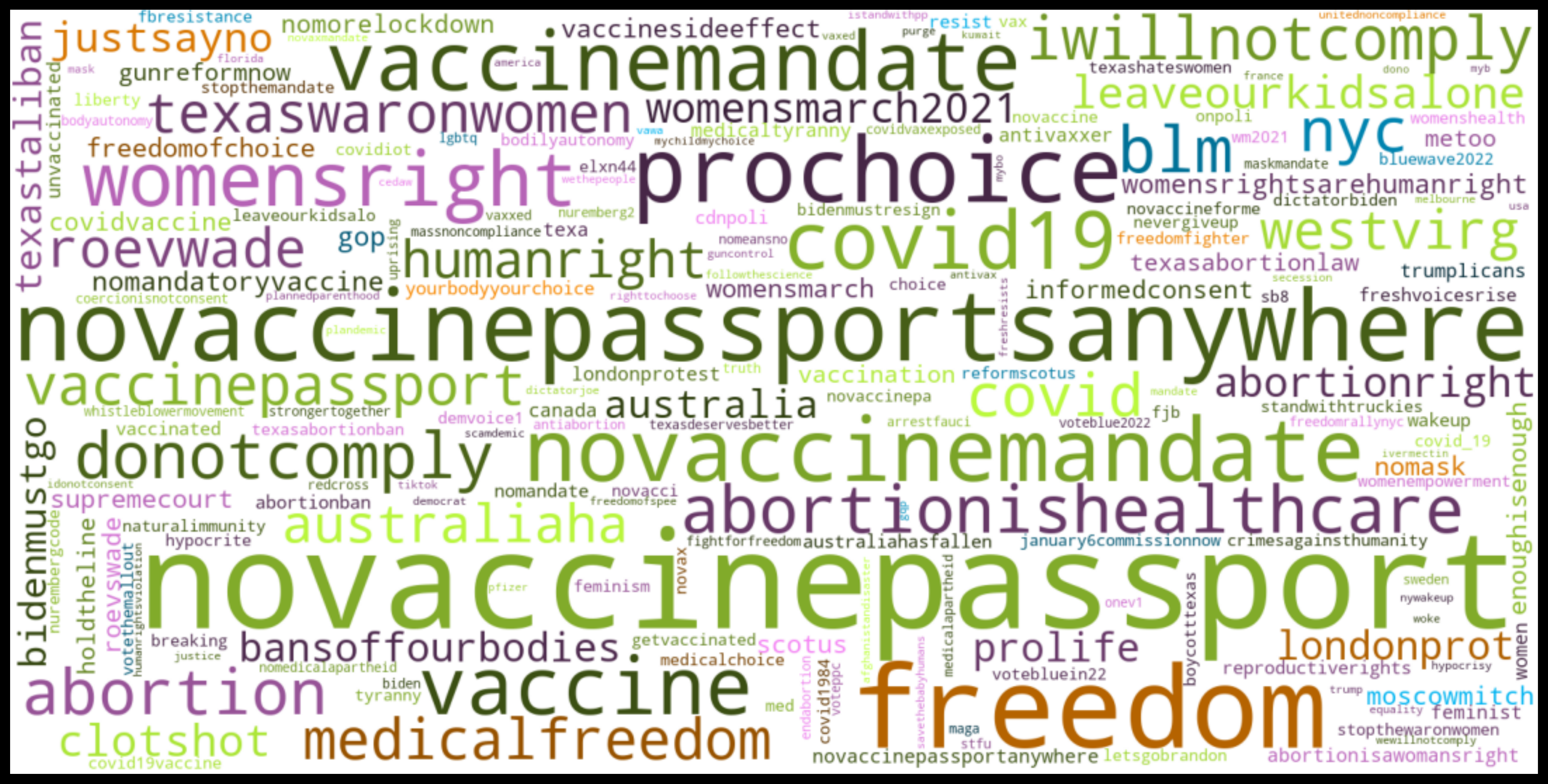}
    \caption{\textbf{Hashtags in our reference windows}. The plots show the most occurring hashtags appearing in tweets from BP, IP, and CC, from left to right. The larger and darker a word is, the more often it appears during the time window. Colors correspond to the respective topics: pink to \emph{women rights \& abortion}, green to \emph{covid \& vaccination}, light blue to  \emph{politics \& civil rights}, and orange to \emph{freedom of choice}.}
    \label{fig:periods_wordclouds}
\end{figure*}

\section{Data}\label{sec:data}
We collected the stream of \mbmc\ tweets from 1-Jan-2018 to 20-Dec-2021.
For the purpose of the analysis, we break down the stream $\stream{\mbmc}$ into three equal-in-length periods to study the content of the hashtag before and after the \covid\ outbreak.
The three reference windows are: (1) Before pandemic (\textbf{BP}) from 1-Jan-2018 to the first officially reported case of \covid\ outside China on 13-Jan-2020. 
(2) Initial pandemic (\textbf{IP}) from 13-Jan-2020 to the beginning of the vaccination campaign with the approval of Moderna vaccine by the FDA on 17-Dec-2020.
(3) \covid\ coexistence (\textbf{CC}) from 17-Dec-2020 to 20-Dec-2021. 

Among the tweets in the stream, we retain the around 280K tweets written in English, corresponding to 86\% of the entire retrieved corpus (excluding tweets from unidentified languages that are too short or contain too many hashtags to infer the language). 
It follows that throughout the analysis we assume that the content of the tweets mostly characterizes the ongoing debates in English-speaking countries. 



\subsection{Data Sub-topics}\label{sec:subtopics}

In this section, we explain how we identify the main sub-topics of the \mbmc\ hashtag, and how we assign a tweet to the respective sub-topic.
Particularly, given the stream $\stream{\mbmc}$, we call a \emph{sub-topic} the topic characterizing the content of a subset of the tweets in the stream.

\textbf{Identify sub-topics.} 
To characterize the presence of sub-topics we use hashtags.
We refrain from the usage of topic-modeling algorithms for two main reasons: (1) In preliminary analysis, we found the hashtags to be representative of the ongoing discussions. 
Indeed, we observed that the text of tweets about either abortion or \covid\ vaccination might contain the same set of words and differ only by hashtags: two examples are ``I have the right to choose \mbmc\ \texttt{\#reproductiverights}'' versus ``I have the right to choose \mbmc\ \texttt{\#novaccine}''.
(2) On Twitter, users can navigate content using hashtags, so we will use them to quantify the ease of moving across sub-topics.
Moreover, we point out that previous work focusing on change of topics also exploited hashtags~\cite{detectingHashjackingtwitter}.

Since 2018, more than 396K unique hashtags (excluding those consisting of a single character) have appeared in $\stream{\mbmc}$.
To find the most relevant hashtags in the stream, we sample 100K tweets from the reference windows \textbf{BP} and the union of \textbf{IP} and \textbf{CC}\footnote{Because the volume of the stream differs across reference windows, we sample with replacement to equally weight the hashtags appearing in different periods. The size of the sample is set to get a reasonable approximation of the real hashtags distribution.}. We take the union of the samples and extract a set $R_i$ of hashtags such that at least 50\% of the tweets in $\stream{\mbmc}$ contains at least one hashtag in $R_i$.
We repeat the sampling 100 times and define the set of hashtags $R = \cup_{i=1}^{100} R_i$ composed of 141  hashtags covering 57.39\% of the entire stream of tweets.


The list $R$ can be grouped into the set $\mathcal{S}$ of five main sub-topics: \emph{women rights \& abortion}, \emph{politics \& civil rights}, \emph{freedom of choice}, \emph{covid}, and \emph{vaccination}.
We note that, for the periods after the pandemic started, we group the \emph{covid} and \emph{vaccination} sub-topics together, since we observed that (1) the two appear often together, and (2) there is no other vaccine debate open during the time in analysis.
We soft-assign\footnote{A hashtag can belong to multiple topics.} the hashtags in the list $R$ to each of these sub-topics.
We refer to the set $\mathcal{T}_{i}$ of hashtags assigned to a sub-topic $i$ as its \emph{signature}.
In~\Cref{tab:hash_signature}, we list some of the hashtags contained in the signature of each sub-topic.

\begin{table}[t]
    \centering
    \resizebox{\columnwidth}{!}{
    \begin{tabular}{lc}
    \toprule
    Sub-topic &  \multicolumn{1}{c}{Hashtags} \\
    \midrule
    \multirow{3}{*}{\emph{Women rights \& abortion}} & \texttt{\#prochoice}, \texttt{\#womensrights}, \\ 
    & \texttt{\#abortion}, \texttt{\#abortionishealthcare},\\
    & \texttt{\#prolife}, \texttt{\#abortionrights}\\
    \midrule
    \multirow{2}{*}{\emph{Politics \& civil rights}} &  \texttt{\#blm}, \texttt{\#blacklivesmatter}, \\
    & \texttt{\#lgbtq}, \texttt{\#votethemout}\\
    \midrule
    \multirow{2}{*}{\emph{Freedom of choice}} &  \texttt{\#freedom}, \texttt{\#medicalfreedom},  \texttt{\#righttochoose}, \\
    & \texttt{\#freedomofchoice}, \texttt{\#bodilyautonomy}\\
    \midrule
    \multirow{2}{*}{\emph{Covid}} &  \texttt{\#covid}, \texttt{\#covid19},  \texttt{\#nomorelockdowns}, \\
    & \texttt{\#freedomofchoice}, \texttt{\#bodilyautonomy} \\
    \midrule
    \multirow{2}{*}{\emph{Vaccination}} &  \texttt{\#vaccine}, \texttt{\#novaccinepassports},  \texttt{\#vaccinesideeffects}, \\
    & \texttt{\#vaccinemandates}, \texttt{\#iwillnotcomply} \\
    \bottomrule
    \end{tabular}}
    \caption{For each sub-topic, we list some of the most relevant hashtags contained in their signature.}
\label{tab:hash_signature}
\end{table}

\textbf{Assign tweets to sub-topics.} The process of assigning tweets to sub-topics consists of two steps. 
In the first unsupervised step, given a tweet $t$, we denote by $R_t$ the list of its hashtags that lie in $R$ and assign $t$ to the sub-topic $i$ via
\begin{equation}\label{eq:assignment}
i = \underset{j \in \mathcal{S}}{\operatorname{argmax}} \  {\sum}_{h \in R_t} \mathbb{P}\left(\mathcal{T}_{j} | h\right) \enspace .
\end{equation}

We apply this rule to all tweets that have at least one hashtag in $R$.
Unfortunately, some tweets may not have any hashtags in $R$.
To address this issue, we use the tweets labeled using~\Cref{eq:assignment} as the ground truth to train a supervised classifier on the text of the remaining set of tweets to assign a tweet to a corresponding sub-topic.
Our classifier consists of a Random Forest~\cite{randomforest} of depth 35 for each time period, which is fed with the text of tweets represented as a count vector\footnote{We evaluate each  classifier on 5-folds (70/30) whose F-1 average is 83.75\% (\textbf{BP}), 87.40\% (\textbf{IP}), and 89.13\% (\textbf{CC}).}. 
The average test F1-score on the sub-topics is 82\% for the BP period, 84\% for the IP period, and 90\% for the CC period.
We decided not to use \emph{covid-twitter-bert}~\cite{twitter-bert2020covid} since a Random Forest classifier is an interpretable and simple model that already yields good results. Interpretability of our model is important since our ground truth is unsupervised, which provides extra flexibility if the labeled data we use require additional cleaning.



\section{Analysis}\label{sec:analysis}

We want to understand whether and how the discourse around the hashtag \mbmc\ has changed after the \covid\ outbreak by answering the following questions.

\textbf{RQ1}. How has the usage of \mbmc\ evolved? 

\textbf{RQ2}. How do user bases and content interactions differ across sub-topics?

\textbf{RQ3}. Has \mbmc\ been hashjacked?

\subsection{RQ1: \mbmc\ Evolution}\label{sec:rq1}

Hashtags function as  means of 
labeling, archiving and distributing information on OSNs. 
As such, we should not underestimate the importance of associating a topic to a hashtag. 
~\mbmc\ was originally created to group content related to the \emph{abortion} debate.
Thus, an average consumer of information on Twitter would likely
expect to find news and opinions about this topic when searching for content associated with the hashtag.

In our attempt to answer RQ1, we investigate if a consumer's experience is likely to be consistent with this expectation, 
 or whether the consumer is more likely to encounter tweets related to
   the online protests  against  \covid\ vaccine mandates in which the  slogan of
   ``My Body My Choice''  is widely used to claim their bodily autonomy.
 


\textbf{\mbmc\ overview.} For an overview of the discourse around~\mbmc, we analyse the hashtags appearing in our three reference windows. 
In~\Cref{fig:periods_wordclouds}, we observe that the hashtags used during the BP period relate more frequently to the discourse centered around women's rights, and 
in particular abortion and 
body autonomy.
During the IP period, women's rights still appear central in the discussion, with a specific focus on asking for justice for women who were subjected to violence, e.g., \texttt{\#justiceforjennifer}, \texttt{\#justiceforuwa}, and \texttt{\#merajismmerimarzi}; at the same time, the hashtag \texttt{\#covid19} starts to gain relevance.
During the CC period, corresponding to the first vaccine approval in the U.S., we see that the conversation about \mbmc\ is no longer centered around a topic associated to women's rights: instead, and in contrast to the periods preceding the \covid\ pandemic, multiple hashtags related to the vaccination campaign become more dominant.

\textbf{Change of most relevant hashtags.}
We sample 100K tweets with replacement for each period BP, IP, and CC. Given one of these periods $T$, let $H$ be the set of the 25 hashtags used most often within the period $T$ and build the vector $\textbf{v}_{H}^{T} \in \mathbb{R}^{25}$ whose entries are co-occurrences of \mbmc\ with each of the 25 hashtags in $H$.
In the time period $E$ following $T$, we count the co-occurrences of \mbmc\ with the same set $H$ of hashtags, obtaining $\textbf{v}_{H}^E \in \mathbb{R}^{25}$.
To understand whether the frequency of co-occurrences of \mbmc\ with the hashtags is independent of the time window, we run a $\chi^2$ test with a significance threshold of $5\%$.
Applying this framework to the periods $T=BP$ and $E=IP$, we observe that the frequency of the hashtags is not independent of the time window. 
For the periods $T=IP$ and $E=CC$, we find similarly that the frequency of hashtags is not independent of the time window according to the $\chi^2$ test.


\begin{tcolorbox}[colback=black!5!white,colframe=black!75!black]
\textbf{Observation 1.} 
The \covid\ outbreak caused a decrease of occurrences of the 25 hashtags used most often before the pandemic. Similarly, the beginning of the vaccination campaign caused a change of occurrences of the hashtags that were most dominant during IC. For instance, the frequency of the hashtags \texttt{\#abortion}, \texttt{\#prochoice}, and \texttt{\#merajismmerimarzi} was reduced by the 48.66\%, 52.82\%, and 98.35\% respectively, while the hashtags \texttt{\#covid19}, \texttt{\#vaccine}, and \texttt{\#freedom} increased in frequency by 97.73\%, 422.66\%, and 690.82\% respectively.
\end{tcolorbox}

\begin{figure}[t]
\includegraphics[width=1\columnwidth]{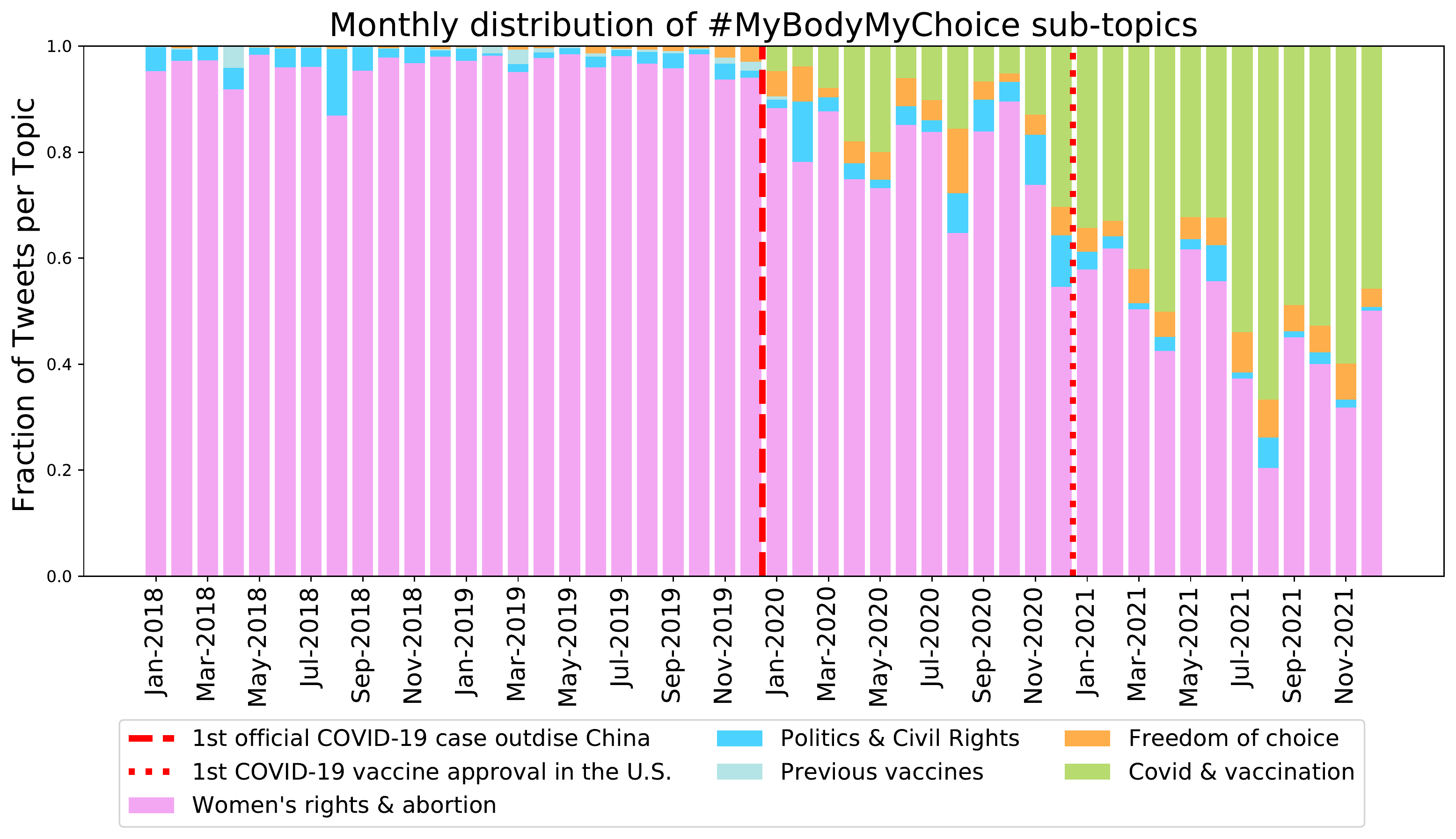}
\caption{\textbf{Sub-Topic Monopolization}: for each sub-topic, we show the monthly fraction of tweets belonging to it.}
\label{fig:distr_num_tweets}
\end{figure}

\textbf{Sub-topics monopolization.} Following the process explained in~\Cref{sec:subtopics}, we separate the tweets into sub-topics.
We refer to the fraction of tweets assigned to a sub-topic as the sub-topic's monopolization factor.
We note that, for the periods after the pandemic started, we grouped together \emph{covid} and \emph{vaccination} sub-topics, since we observed that (1) the two appear often together, and (2) there is no other vaccine debate open during the time in analysis. 
~\Cref{fig:distr_num_tweets} shows the monthly monopolization factor of each of the sub-topics. 
Consistently with our observations of the most dominant hashtags shown in~\Cref{fig:periods_wordclouds}, the fraction of tweets about \emph{women's right \& abortion} starts to decrease after the first official \covid\ case was registered outside China, and the start of the vaccination campaign further reduces the original sub-topic of the hashtag \mbmc\  with the introduction of hashtags such as \texttt{\#novaccinepassport}, \texttt{\#novaccinemandate}, and \texttt{\#dontcomply}.
In fact, during the second half of 2021, the tweets about \emph{women's right \& abortion} represent less than 40\% of the total number of tweets.
Note, however, that the monopolization factor of \emph{covid} is reduced after September 2021 when the abortion ban in Texas went into effect.

\begin{tcolorbox}[colback=black!5!white,colframe=black!75!black]
  \textbf{Observation 2.} 
  After the \covid\ outbreak, the usage of the hashtag \mbmc\ changed on Twitter, with the conversation drifting towards  \covid.
  In particular, the most relevant hashtags after the outbreak reflect opinions opposing the current vaccination campaign, typical of antivaxxers. 
\end{tcolorbox}

\begin{figure}[t]
\includegraphics[width=0.6\columnwidth]{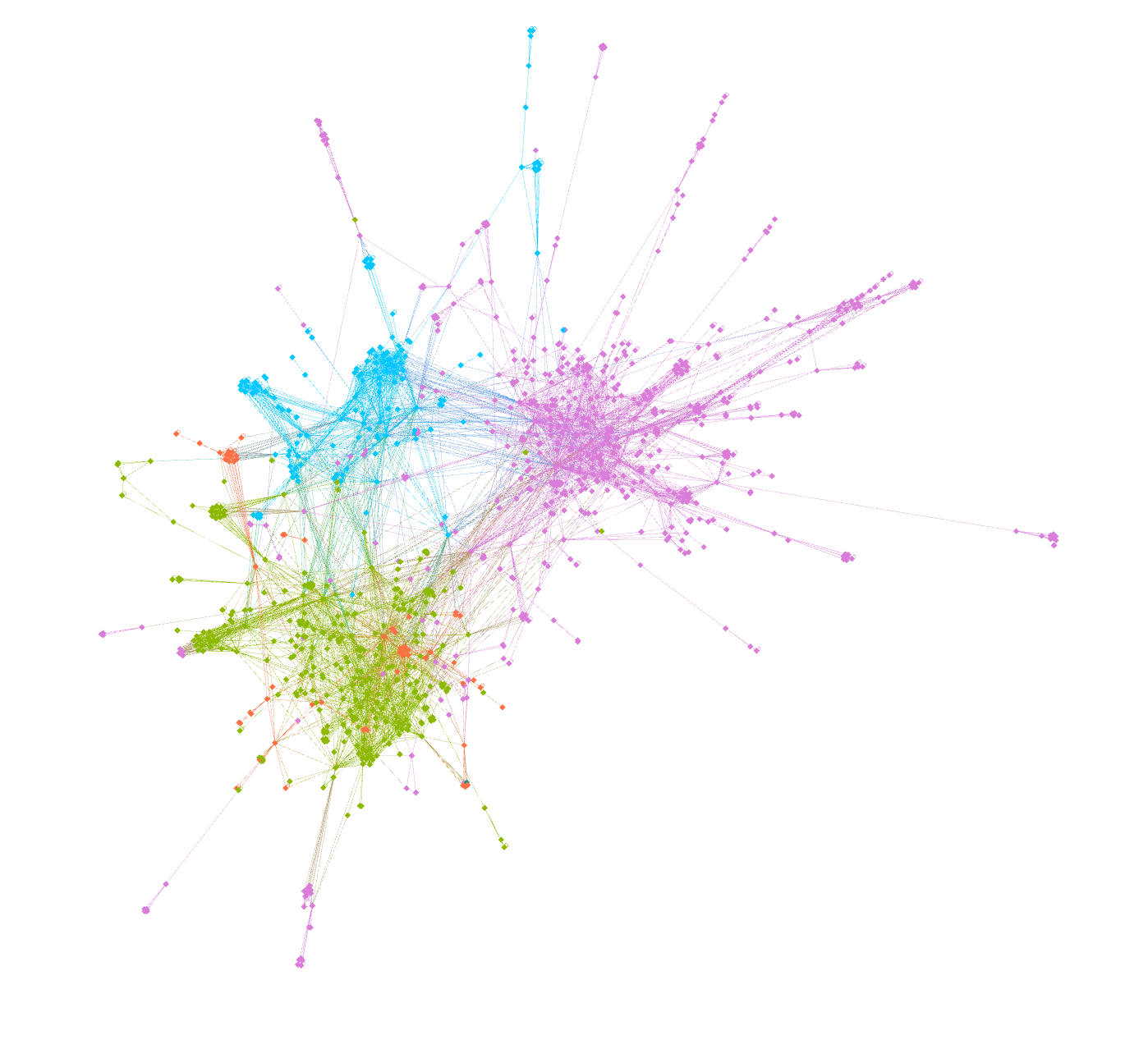}
\caption{Network of hashtags in the \mbmc\ stream. We draw the largest connected components of the network. Pink nodes refer to the \emph{women rights \& abortion} sub-topic, green nodes to \emph{covid} and \emph{vaccination}, blue nodes to  \emph{politics \& civil rights}, and the orange nodes to \emph{freedom of choice}.}
\label{fig:hash_graph}
\end{figure}

\textbf{Sub-topics separation.} 
So far, we have investigated how the original topic of \emph{abortion} in the \mbmc\ stream was affected by the \covid\ breakout. 
Next, we want to understand how connected or separated the four sub-topics are from each other.
To this end, we built a hashtag network, where each node is a hashtag, and two hashtags are connected by an edge if they co-occur in at least 30 tweets.
Each edge is weighted by the number of co-occurrences.
~\Cref{fig:hash_graph} shows a planar embedding of the resulting network in which nodes are colored by sub-topic. The figure indicates that each sub-topic corresponds to a separate set of hashtags.
To quantify this phenomenon, we employ the Random-Walk Controversy score (RWC)~\cite{GarimellaMGM18}.
The RWC score measures the degree of separation between two sets of nodes in a graph by comparing the probabilities of moving within and between the two sets.
A negative score indicates that the two sets of nodes are not separated well so that the number of connections within and across the sets are comparable. 
When the RWC score is larger than zero, the two sets are well-separated, and there are likely very few connections between them relative to the number of connections within them.
In our framework, the RWC score quantifies the overall separation of sub-topics by measuring the frequency with which hashtags  associated with different sub-topics co-occur. 
On Twitter, hashtags are hyperlinks that allow users to navigate the associated content.
Thus, in this context, RWC scores reflect how much users who navigate the content of a sub-topic are exposed to links related to content about other sub-topics. 
In particular, a lower RWC score implies that users who explore a sub-topic are less likely to reach content about the other sub-topic if they are navigating Twitter by following hashtags.

\begin{table}[t]
    \centering
    \resizebox{\columnwidth}{!}{
    \begin{tabular}{lcccc}
    \toprule
    Sub-topic &  \emph{Women's rights \& abortion} & \emph{Politics \& civil rights} & \emph{Freedom of choice} & \emph{Covid \& Vaccination} \\
    \midrule
    \emph{Women rights \& abortion} & - & 0.49 & 0.32 & 0.46 \\
    \midrule
    \emph{Politics \& civil rights} & 0.49 & - & 0.31 & 0.48 \\
    \midrule
    \emph{Freedom of choice} & 0.32 & 0.31 & - & 0.11 \\
    \midrule
    \emph{Covid \& Vaccination} & 0.46 & 0.48 & 0.11 & - \\
    \bottomrule
    \end{tabular}
    }
    \caption{For each pair of sub-topics, we report the RWC score. The closer the score is to one, the more separated the two sub-topics are. Note that the RWC score is symmetric.}
\label{tab:rwc}
\end{table}
~\Cref{tab:rwc} reports the RWC scores for each pair of sub-topics. 
The \emph{Women's rights \& abortion} and the \emph{covid \& vaccination} sub-topics are the most separated, thus illustrating the topics shift of \mbmc.
In contrast, \emph{covid \& vaccination} is close to \emph{freedom of choice}, which, in turn, is close to \emph{politics \& civil rights}.

\begin{tcolorbox}[colback=black!5!white,colframe=black!75!black]
  \textbf{Observation 3.} 
  The \emph{women's rights \& abortion} and \emph{covid \& vaccination} sub-topics are well-separated. In particular, we expect that the hashtags characterizing the first sub-topic rarely co-occur with the hashtags representing the second sub-topic. Users who access content related to one of the two sub-topics therefore have a smaller probability of seeing hyperlinks to content related to the other sub-topic than to content of the same sub-topic.
\end{tcolorbox}

\begin{figure*}[t]
\includegraphics[width=1\columnwidth]{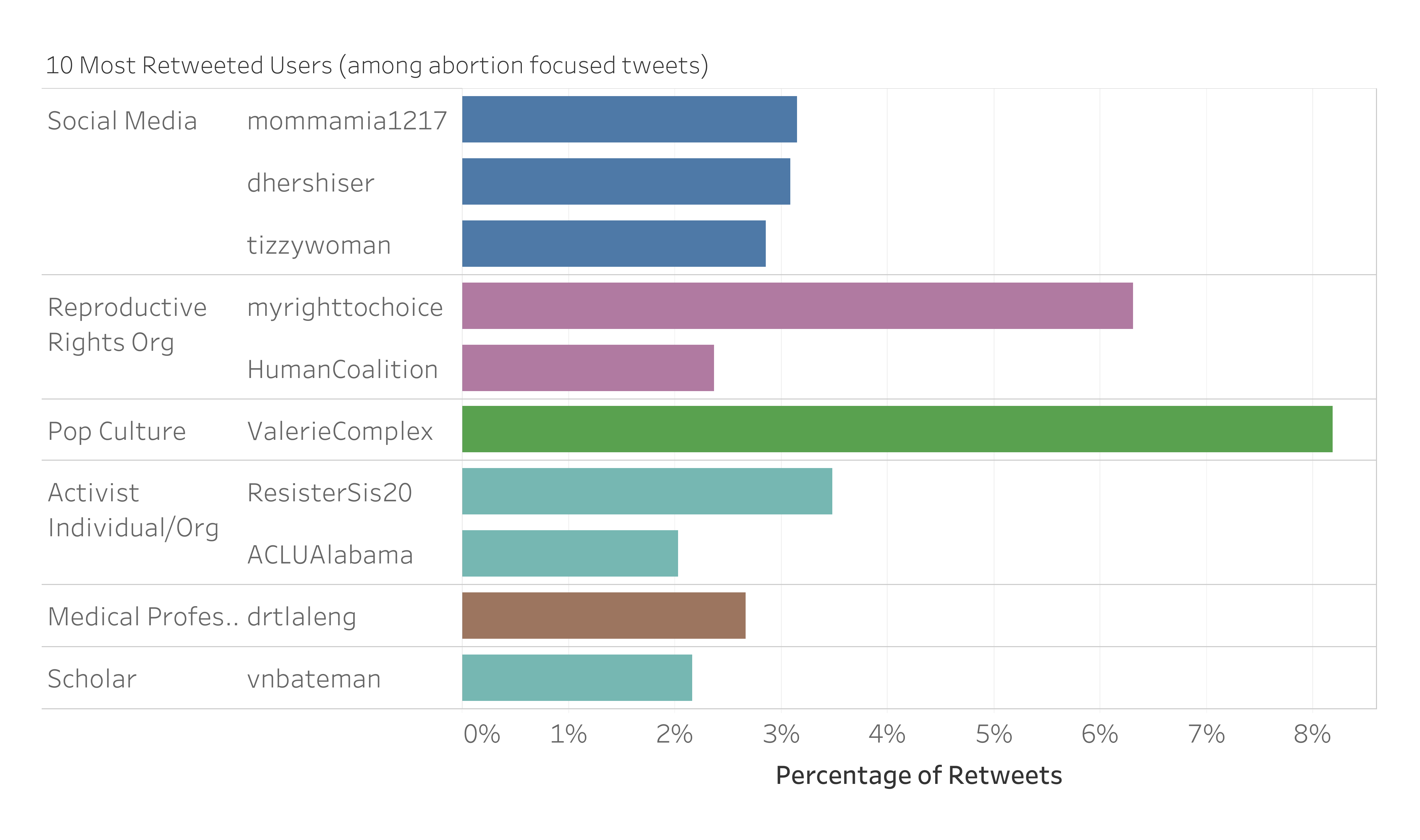}
\includegraphics[width=1\columnwidth]{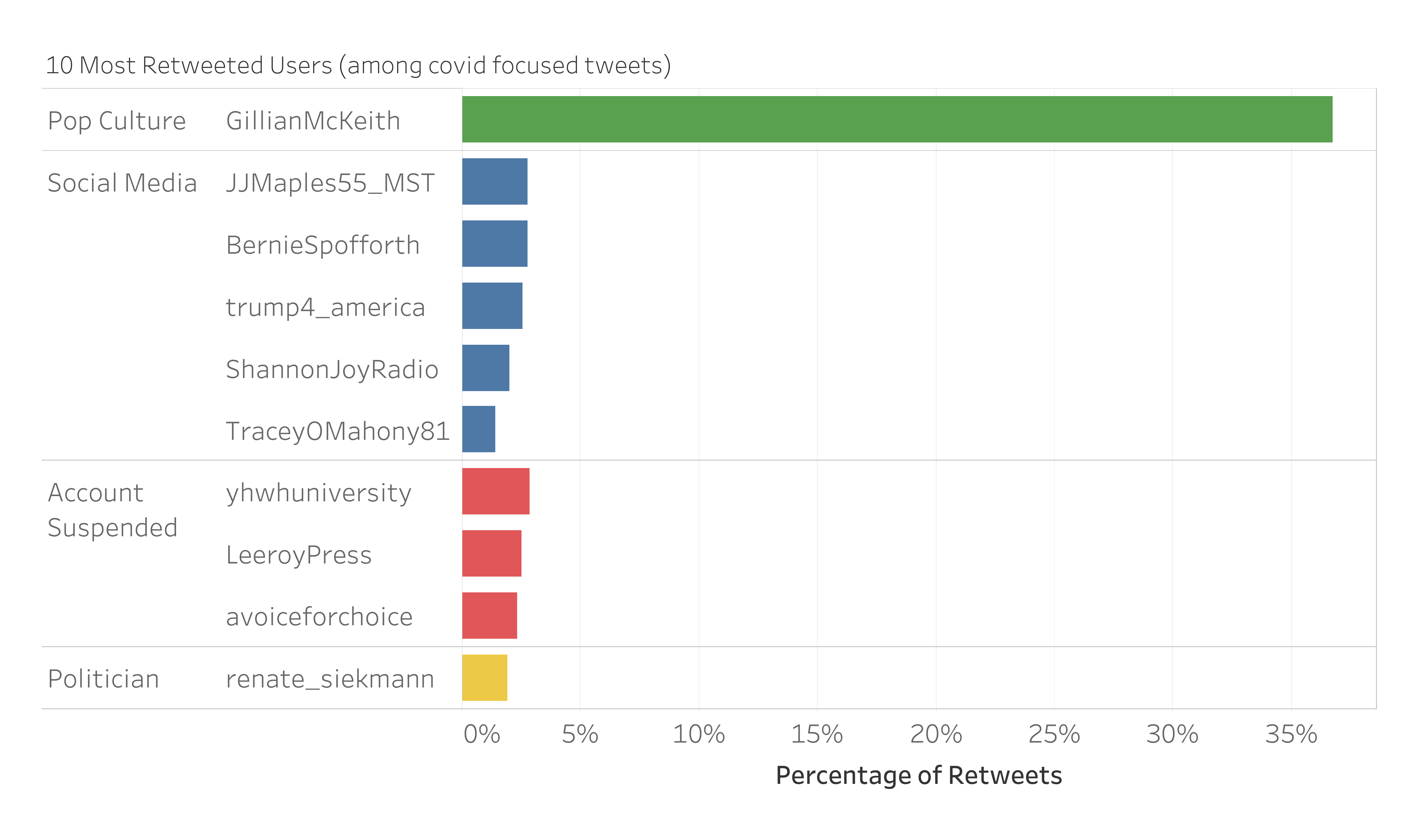}
\caption{A list of the ten most retweeted users within the \emph{women's rights \& abortion} and the \emph{covid \& vaccination} sub-topics. Percentages are computed with respect to the total number of retweets.}
\label{fig:rt}
\end{figure*}

\subsection{RQ2: Compare Sub-Topics}


The analysis in the previous section showed that \covid\ has become a prominent sub-topic of the stream of tweets tied to \mbmc.
In this section, our goal is to understand how the sub-topics related to \covid\ and abortion differ from each other in terms of their user bases and the popularity of content.
We consider the tweets posted during the IP and CC periods, because the \emph{covid \& vaccination} sub-topic pop up after the beginning of the pandemic.
The number of unique users that posted content about \emph{women's rights \& abortion} is 1.82 times the number of users tweeting about \emph{\covid\ \& vaccines}. 
Furthermore, 90.7\% of users tweeted about only one of the two topics.
Thus, we conclude that the user base of the two branches of \mbmc\ are separated.

Next, we investigate whether the number of posts, retweets, replies, and quotes, which constitute the different ways in which tweets are generated, differ between these two sub-topics and find that there are no significant differences:
For the sub-topic \emph{women's rights \& abortion},  56.71\% of tweets are retweets, 18.0\% are replies, 7.34\% are quotes, and the remaining are posts; 
the tweets that generate the retweets constitute 11.85\% of the total tweets in this sub-topic.
In the corpus of \emph{covid \& vaccination}, 55.94\% of tweets are retweets, 20.92\% are replies, 6.38\% are quotes, and the remaining are posts;
the tweets that generate the retweets make up 12.51\% of all tweets in this sub-topic.
Finally, we observe that the distributions of the number of tweets per user is comparable for both sub-topics: the median is one with few outliers in both populations. 
We speculate that the similar distribution across the sub-topics indicates that users tweeting about \emph{covid \& vaccination} were not part of an organized effort to intentionally shift the use of the hashtag \mbmc.

\begin{tcolorbox}[colback=black!5!white,colframe=black!75!black]
 \textbf{Observation 4.} 
 The user bases of \emph{women's rights \& abortion} and \emph{covid \& vaccination} are different and display a similar characterization of content interaction (i.e., posts, retweets, replies, and quotes are distributed similarly). This indicates that there was no organized intervention to intentionally create a shift of the \mbmc\ discourse.
\end{tcolorbox}

\begin{figure*}[t]
\includegraphics[width=1\linewidth]{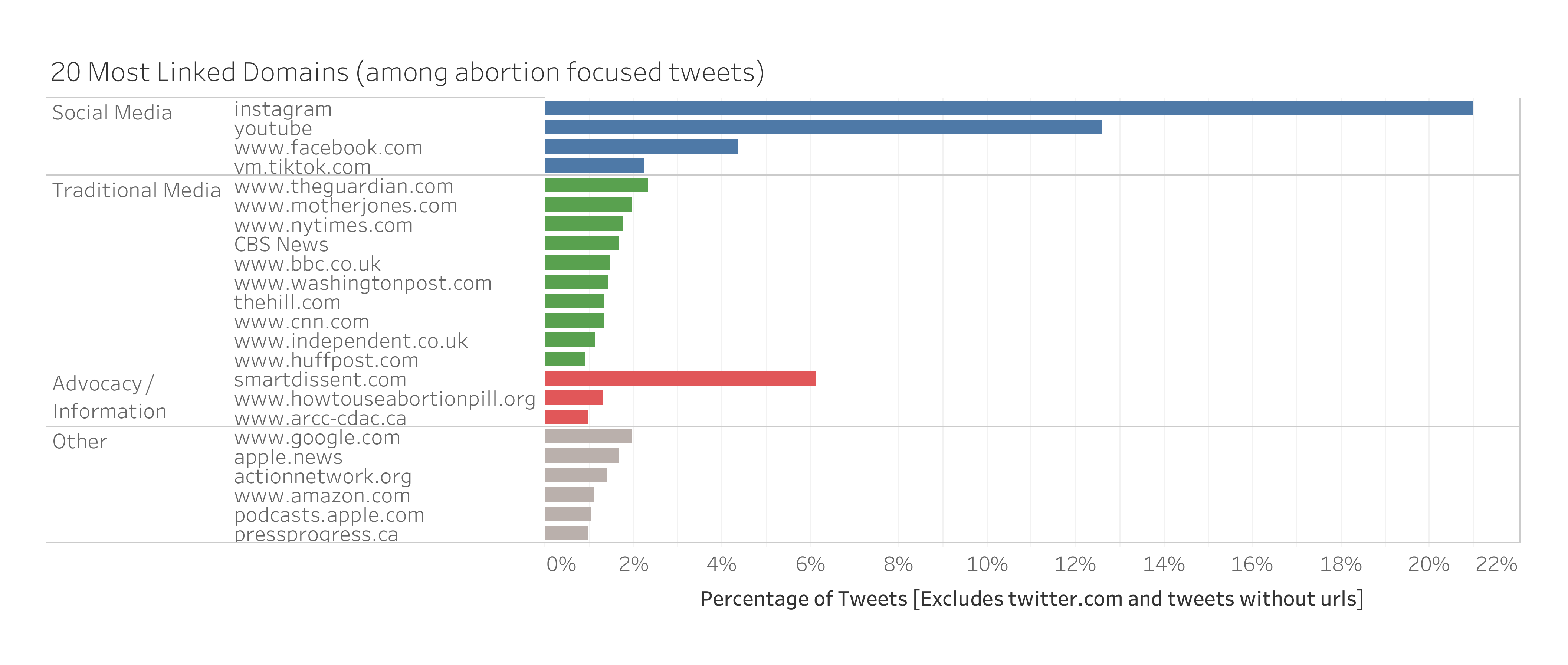}
\includegraphics[width=1\linewidth]{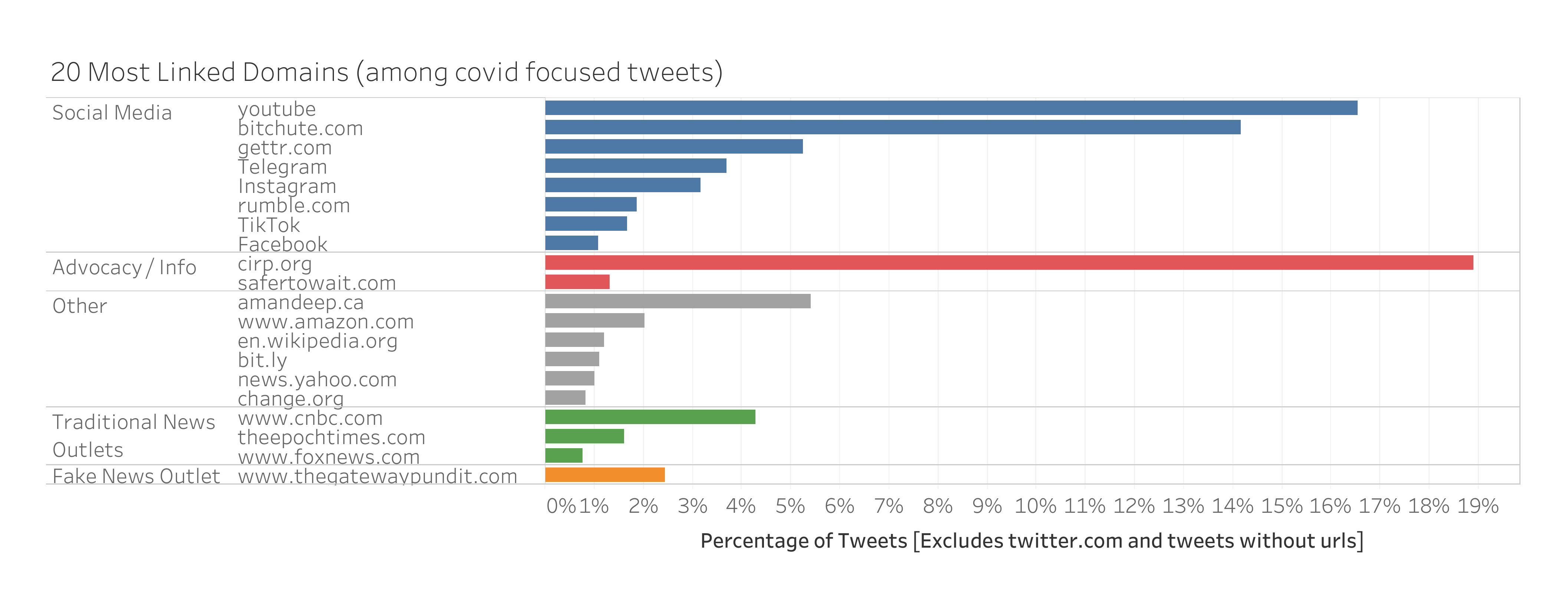}
\caption{Listed are the domains of the most frequent URLs attached to tweets for the \emph{women's rights \& abortion} (top) and \emph{covid \& vaccination} (bottom) sub-topics.
The percentages are computed over the total number of tweets that contain a non-Twitter URL. The percentage of tweets containing non-Twitter URLs is 33.1\% for  \emph{women's rights \& abortion} and 25.08\% for \emph{covid \& vaccination}.}
\label{fig:urls}
\end{figure*}

In~\Cref{fig:rt}, we report the top-10 retweeted users appearing in tweets about \emph{women's rights \& abortion} and \emph{covid \& vaccination}, respectively.
The group of most frequently retweeted users focused on the topic of \emph{women’s rights \& abortion} are divided fairly evenly between a few categories of users.
These include activist and reproductive rights organizations, individuals from film, medical, and academic circles, and users known primarily for their social media presence. 
In contrast, the group of most retweeted users focused on the topic of \emph{covid \& vaccinations} includes more users from the social media category, fewer authority figures, and no organizations. In the latter sub-topic, there is also a striking gap between the most influential user's share of retweets and everyone else's: the user \texttt{@GillianMcKeith} has 37\% of the retweets in this category, while the rest of the top ten have 1-3\% each. Furthermore, three of the top ten accounts in the second sub-topic have been suspended.


\begin{tcolorbox}[colback=black!5!white,colframe=black!75!black]
 \textbf{Observation 5.} 
 The majority of the most retweeted users of \emph{women's rights \& abortion} are organizations, activists, or professionals. In contrast, the set of accounts retweeted by the user base of  \emph{covid \& vaccination} suggests that it gives more prominence to users that individually select, create, and share content without the support of 
 established 
 organizations.
\end{tcolorbox}

In~\Cref{fig:urls}, we report the URLs appearing in tweets from  the \emph{women’s rights \& abortion} and \emph{covid \& vaccinations} sub-topics.
The domains linked to in tweets from the \emph{women’s rights \& abortion} subcategory come primarily from three categories of sources; social media, traditional news outlets, and websites advocating for a position or providing information. 
The group of top domains linked to in tweets from the \emph{covid \& vaccinations} subcategory adds four additional social media sites to the four used by the abortion subcategory. 
These additional sites are known for having less restrictive censorship policies. 
Advocacy groups have a larger share of the links amongst the \emph{covid and vaccination} subgroup, and they are a different set of organizations than those linked to by the women’s rights and abortion subcategory. 
Finally, traditional news outlets make up a much smaller portion of the top domains in the \emph{covid and vaccinations} subcategory, and again the specific outlets are very different. 
In addition, one outlet known to disseminate ‘fake news’ is amongst the top twenty.





\subsection{RQ3: Has \mbmc\ been hijacked?}\label{sec:tax}
There are a variety of different types of hashjacking, and this section situates the case of \mbmc\ in the context of other well-known hashjackings based on the analytical observations drawn so far. There are largely three types of hashjackings: those that semantically change the associated tweets, those that intentionally make a topic more polarized, and those that capitalize on the success of a given hashtag without taking part in the original conversation. These three categories are outlined below: 

\textbf{1) Hashjacking that semantically changes the associated tweets (whether intentional or not).}
When Alyssa Milano, in 2017, revitalized the \#\texttt{Metoo} movement on Twitter by encouraging thousands of women to share their stories of sexual harassment, the movement garnered widespread global attention that did not attribute its founder, Tarana Burke, or use the language of the movement within its intended context of the Black community.
As a result, \#\texttt{Metoo} turned from a movement focused on Burke's community to women's experiences with sexual harassment broadly~\cite{timesmetoo}. 
While the women responding to Milano's encouragement to post their own stories were not necessarily intentionally hashjacking Burke's words for the Black community, the result was that the semantic meaning of the hashtag and its associations changed.

\textbf{2) Hashjacking that intentionally makes a topic more polarized.} 
Within this category of hashjackings, users appropriate a hashtag for the opposite purpose of the hashtag's originally intended use. For example, McDonald's launched a Twitter campaign using the hashtag \#\texttt{McDStories} to encourage users to share their fond stories of Happy Meals. The hashtag took off as users posted critical comments about the food and how it impacted their health negatively. A similar example is the hashtag \#\texttt{Februdairy}, which was launched by the diary industry to support the dairy market after the creation of Veganuary, a challenge to join a global community to try vegan food for a month: users repurposed the hashtag to write about problems in the dairy industry.

\textbf{3) Hashjacking to capitalize on the success of a given hashtag without taking part in the conversation.}
After \#\texttt{Metoo} was popularized by Alyssa Milano, the hashtag was then also hashjacked in 2018 by the '120 decibels' campaign by the Austrian Identitarian Movement in an attempt to latch onto to prominence of the  \#\texttt{Metoo} hashtag~\cite{decibels}. Similarly, the hashtag \#\texttt{Brazil2014} was  hashjacked by ISIS in order to entice fans watching the World Cup in 2014 to click on links to the group's propaganda videos~\cite{isis}. In these hashjackings, the hashjackers did not intend to engage with the topic of the hashtag; rather they were solely using the hashtag to capitalize on its success. 


\begin{tcolorbox}[colback=black!5!white,colframe=black!75!black]
  \textbf{Observation 6.} \#\texttt{MyBodyMyChoice} was a hashtag rooted in women's rights and was then used by users during the pandemic to refer to something else entirely: anti-vaccine content. We situate \mbmc\ within the semantic hashjacking as (1) the discourse around the hashtag changed over time, and users who were hashjacking are distributed unique users rather than a given organization (Observations~1, 2, 4, \& 5); (2) Users who hashjacked \#\texttt{MyBodyMyChoice} used it to summarize their own positions on vaccines and masks rather than capitalize on the hashtag's success. Furthermore, the hashtag was not hashjacked to make women's reproductive rights more polarized but rather to express a different argument against COVID-19 vaccines (Observation~3).
\end{tcolorbox}




\section{Related Work}\label{sec:related}

We contextualize our work by providing a brief overview of previous studies about hashjacking.
Some of these works analyse whether a  hashtag has changed its usage (i.e., hashjacked), while some concentrate on developing tools to identify semantic shifts. 

Past works emphasized the change of usage of a hashtag due to an event or by group of users.
~\citet{Rodriguez2020} showed that \texttt{\#FamilyTravelHacks}, created to inform people about tips for safe travel with children, changed its usage after Executive Order 13769, 
a.k.a., the 
travel ban.
As described in~\Cref{sec:tax},~\citet{decibels} demonstrated that the hashtag \texttt{\#metoo} was hijacked by the ‘120 decibels’ campaign.
Some other works analyzed how polarized users intentionally \emph{hash-jack} a tweet to express their views contrasting the opinion of the original user base of the hashtag, e.g., in Germany's presidential elections~\cite{darius2019hashjacking} or in U.S. politics~\cite{Garimella2013}. 

To analyse these phenomena, several methodological works focus on designing computational tools to identify changes in the linguistic usage of a word~\cite{langchange,langchangeWest2013no,langchange2} or of the meaning of a hashtag on Twitter~\cite{detectingHashjackingtwitter}.
Particularly,~\citet{detectingHashjackingtwitter} use a matrix factorization-based method to build a day-by-day topic matrix and  use a standard change-detection hypothesis test to see if a change in the topic occurred. 
While they provide a strategy to say ``if'' the topic of a hashtag has changed or not, we focus on ``how'' the meaning of \mbmc\ has changed.
For this reason, we adopted a qualitative approach instead of their method.

\section{Conclusion}
In this paper, we examined the hashtag \mbmc\ to see how the content of  tweets associated with it evolved during the pandemic. The hashtag originated to support women's rights around abortion and bodily autonomy, but we observed that during the pandemic a separate set of users utilized it to communicate their disagreement with vaccines. 

We analysed the change of usage of \mbmc\ considering three factors: 
(1) the semantic of the discourse around the hashtag; (2) the users' interactions with contents related to diverse sub-topics; and (3) the users' attention toward specific users and external sources of information.
We demonstrate the hypothesized semantic shift by employing 
supervised learning and statistical analysis.
Furthermore, we quantify the shift by showing the evolution of dominant sub-topics.
To understand whether the semantic shift was intentionally induced,
we conducted a user behaviour analysis and found no evidence for such organized attempts.  
Finally our analysis of users' attention to highly retweeted users and most commonly embedded URLs shows that attention has shifted from experts and authorities to individuals and that some unreliable sources appeared during the \covid\ debate. We note that our analysis focused only on the stream of tweets using \mbmc\ and does therefore not necessarily  generalize to the entire discourse of these topics. 
Moreover, although the methodology we used can be applied to other topics, it would still require human in the loop to first distinguish among sub-topics.

\textbf{Disclaimer.} It is important to note that the demographics of Twitter users is not representative of the overall population: 40\% of Twitter users are in the 25-34 age group,  another 20\% are in the 35-49 age group, and the age groups 18-24 and above 50 each constitute around 17\% of Twitter users (Statista); furthermore, 63\% of Twitter users are male.
Thus, as our study explores how the content associated with the \mbmc\ hashtag changed over the pandemic, it is important to keep in mind that our results may have been impacted by the demographics on Twitter. 


\section{Acknowledgments}

This work was a result of collaboration between Brown University's Data Science Initiative, Brown University's Computer Science Department and the Center for Digital Scholarship at Brown University. 
Haddadan was partially supported by the Data Science Initiative at Brown University and by the NSF via grant CCF-1740741.
Ramachandran acknowledges support from US National Institutes of Health R01 GM118652 and R35 GM139628.
Sandstede acknowledges support from the NSF grant CCF-1740741.

\bibliographystyle{ACM-Reference-Format}
\bibliography{biblio}

\end{document}